\documentclass{article}
\usepackage{amssymb}

\begin{document}

\title{Destruction of long-range antiferromagnetic order by hole doping}
\author{F. Carvalho Dias,$^{\dagger }$ I. R. Pimentel and R. Orbach$^{*}$
  \\ {\small Department of Physics and  CFMC, University of
  Lisbon, 1649 Lisboa, Portugal} \\ $^{*}${\small Department of Physics,
  University of California, Riverside, CA92521, USA}}
\date{}
\maketitle

\begin{abstract}
We study the renormalization of the staggered magnetization of a
two-dimensional antiferromagnet as a function of hole doping, in the
framework of the t-J model. It is shown that the motion of holes 
generates decay of spin waves into ''particle-hole'' pairs, which causes 
the destruction of the long-range magnetic order at a small hole 
concentration. This effect is mainly determined by the coherent motion 
of holes. The value obtained for the critical hole concentration, of 
a few percent, is consistent with experimental data for 
the doped copper oxide high-$T_{c}$ superconductors.

\noindent PACS: 74.25.Ha, 75.40.Cx, 75.50.Ep

\end{abstract}

\newpage

One of the interesting features of the copper oxide high-$T_{c}$
superconductors is the dramatic reduction, with doping, of the long-range
magnetic order of their parent compounds.$^{1}$ The undoped materials, e.g.
$La_{2}CuO_{4}$, are antiferromagnetic (AF) insulators. Doping, e.g., in
$La_{2-\delta }Sr_{\delta }CuO_{4}$, introduces holes in the spin lattice 
of the $CuO_{2}$ planes, and the long-range AF order is destroyed at a 
small hole concentration, $\delta _{c}\sim 0.02$. The $CuO_{2}$ planes are
described by a spin-1/2 Heisenberg antiferromagnet on a square lattice, 
with moving holes that strongly interact with the spin array. 
The motion of holes generates spin fluctuations that tend to disrupt 
the AF order. It has been shown that hole motion produces strong effects 
on the magnetic properties, leading in particular to significant softening 
and damping of the spin excitations as a function of doping.$^{2-5}$ 
The critical concentration $\delta _{c}$ where long-range magnetic order 
disappears has often been identified with the concentration where the spin 
wave velocity vanishes.
However important damping effects occur, which have to be taken into
account. In particular, all spin waves become overdamped at a concentration
well below the one for which the spin wave velocity vanishes, suggesting
that the long-range AF order may disappear at a smaller concentration.$^{5}$
The critical hole concentration $\delta _{c}$ is provided by the
vanishing of the staggered magnetization order parameter.

In this work we use the t-J model to calculate the doping dependence of the
staggered magnetization of a two-dimensional antiferromagnet, and determine
the critical hole concentration $\delta _{c}$. It is shown that the motion
of holes generates decay of spin waves into ''particle-hole'' pairs, leading
to broadening of the spin-wave spectral function. This broadening gives rise
to a drastic reduction of the staggered magnetization and the disappearance
of the long-range order at low doping, in agreement with experiments.
Such a process was suggested some years ago by Ramakrishnan.$^{6}$
The vanishing of the staggered magnetization as a consequence of doping, has
already been studied in the t-J model by Gan and Mila,$^{7}$ considering
the scattering of spins by moving holes, and by Khaliullin and Horsch$,^{8}$
considering spin disorder introduced by the incoherent motion of holes.

We describe the copper oxide planes with the t-J model,

\begin{equation}
H_{t-J}=-t\sum_{<i,j>,\sigma }\left( c_{i\sigma }^{\dagger
}c_{j\sigma }+H.c.\right) +J\sum_{<i,j>}\left( \mathbf{S}_{i}\cdot
\mathbf{S}_{j}-\frac{1}{4}n_{i}n_{j}\right) ,  \label{(1)}
\end{equation}

\noindent where, $\mathbf{S}_{i}=\frac{1}{2}c_{i\alpha }^{\dagger }
\mathbf{\sigma }_{\alpha \beta }c_{i\beta }$ is the electronic spin operator,
$\mathbf{\sigma}$ are the Pauli matrices, $n_{i}=n_{i\uparrow}+n_{i\downarrow}$
 and $n_{i\sigma }=c_{i\sigma }^{\dagger }c_{i\sigma }$.
To enforce no double occupancy of sites, we use the slave-fermion Schwinger
Boson representation for the electron operators $c_{i\sigma }=f_{i}^{\dagger}
b_{i\sigma }$, where the slave-fermion operator $f_{i}^{\dagger }$ creates
a hole and the boson operator $b_{i\sigma }$ accounts for the spin, subject
to the local constraint $f_{i}^{\dagger }f_{i}+\sum_{\sigma }
b_{i\sigma}^{\dagger }b_{i\sigma }=2S$. For zero doping, the model $(1)$
reduces to a spin-1/2 Heisenberg antiferromagnet, exhibiting long-range
N\'{e}el order at zero temperature. The N\'{e}el state is represented by a
condensate of Bose fields $b_{i\uparrow }=\sqrt{2S}$ and
$b_{j\downarrow }=\sqrt{2S}$, respectively in the up and down sub-lattices,
and the bosons $b_{i\downarrow}=b_{i}$ and $b_{j\uparrow }=b_{j}$ are then
spin-wave operators on the N\'{e}el background. After Bogoliubov-Valatin
transformation on the boson Fourier transform $b_{\mathbf{k}}=v_{\mathbf{k}}
\beta _{-\mathbf{k}}^{\dagger }+u_{\mathbf{k}}\beta _{\mathbf{k}}$, with
$u_{\mathbf{k}}=\left[ \left((1-\gamma _{\mathbf{k}}^{2})^{-1/2}+1\right)
/2\right]^{1/2}$, $v_{\mathbf{k}}=-\mathrm{sgn}(\gamma _{\mathbf{k}})
\left[ \left( (1-\gamma _{\mathbf{k}}^{2})^{-1/2}-1\right) /2\right] ^{1/2}$,
and $\gamma _{\mathbf{k}}=\frac{1}{2}\left( \cos k_{x}+\cos k_{y}\right) $,
we arrive at the effective Hamiltonian

\begin{equation}
H=-\frac{1}{\sqrt{N}}\sum_{\mathbf{q},\mathbf{k}}f_{\mathbf{q}}f_{\mathbf{q}
-\mathbf{k}}^{\dagger }\left[ V(\mathbf{q},-\mathbf{k})\beta _{-\mathbf{k}}+
V(\mathbf{q-k},\mathbf{k})\beta _{\mathbf{k}}^{\dagger }\right] +
\sum_{\mathbf{k}}\omega _{\mathbf{k}}^{0}\beta _{\mathbf{k}}^{\dagger }
\beta _{\mathbf{k}},
\label{(2)}
\end{equation}

\noindent having $S=1/2$ and $N$ sites in each sub-lattice. In $(2)$, the
first term, with
$V(\mathbf{q},\mathbf{k})=zt\left( \gamma _{\mathbf{q}}u_{\mathbf{k}}+
\gamma_{\mathbf{q}+\mathbf{k}}v_{\mathbf{k}}\right) $, represents the
interaction between holes and spin waves resulting from the motion of holes
with emission and absorption of spin waves, and the second term describes 
spin waves for a pure antiferromagnet, with dispersion 
$\omega _{\mathbf{k}}^{0}=
(zJ/2)\left( 1-\gamma _{\mathbf{k}}^{2}\right) ^{1/2}$, $z$ being the 
lattice coordination number ($z=4$).

The staggered magnetization is given by

\begin{equation}
M=<S_{\uparrow }^{z}>-<S_{\downarrow }^{z}>=2<S_{\uparrow }^{z}>,
\label{(3)}
\end{equation}
with
\[
<S_{\uparrow }^{z}>=\sum_{i\in S(\uparrow )}<S_{i}^{z}>, 
\]
where the sum is over the up sub-lattice. Using the Schwinger boson
representation for the spin operator $S_{i}^{z}
=\frac{1}{2}(c_{i\uparrow }^{\dagger }c_{i\uparrow }-c_{i\downarrow}^{\dagger}
c_{i\downarrow })$, and the boson condensation associated to the N\'{e}el
state, one has $S_{i}^{z}=(1-h_{i}^{\dagger}h_{i})(S-b_{i}^{\dagger }b_{i})$,
which, after Bogoliubov-Valatin transformation, leads to

\begin{equation}
M=(1-\delta )\left[ M_{0}-\Delta M\right] ,  \label{(4)}
\end{equation}
where
\begin{equation}
M_{0}=2\left[ NS-\sum_{\mathbf{k}}v_{\mathbf{k}}^{2}\right]
\label{(5)}
\end{equation}
is the staggered magnetization for a pure antiferromagnet, and
\begin{eqnarray}
\Delta M &=&2\sum_{\mathbf{k}}\left[ (u_{\mathbf{k}}^{2}+v_{\mathbf{k}}^{2})
<\beta _{-\mathbf{k}}^{\dagger }\beta _{-\mathbf{k}}>\right.
\nonumber 
\\
&&\left. \qquad +u_{\mathbf{k}}v_{\mathbf{k}}(<\beta _{\mathbf{k}}
\beta _{-\mathbf{k}}>+<\beta _{-\mathbf{k}}^{\dagger }
\beta _{\mathbf{k}}^{\dagger}>)\right] .  \label{(6)}
\end{eqnarray}
The prefactor in $\left( 4\right) $ accounts for the spin dilution
due to doping, being negligible for small hole concentrations. In $(5)$, the
order parameter is considerably reduced by quantum fluctuations, to
$\simeq0.6\times 2NS$. With zero doping the expectation values in $(6)$ vanish and
$\Delta M=0$. However, in a doped system the motion of holes generates spin
fluctuations, giving rise to nonzero expectation values in $(6)$, even at
zero temperature, and then $\Delta M\neq 0$.

In order to calculate the staggered magnetization for the doped system, we
need the spin-wave Green's functions, defined as
$D^{-+}(\mathbf{k},t-t^{\prime})=
-i\left\langle \mathcal{T}\beta _{\mathbf{k}}(t)\beta _{\mathbf{k}}^{\dagger }
(t^{\prime })\right\rangle $, $D^{+-}(\mathbf{k},t-t^{\prime})=-i\left\langle
\mathcal{T}\beta _{-\mathbf{k}}^{\dagger }(t)\beta _{-\mathbf{k}}(t^{\prime })
\right\rangle $, $D^{--}(\mathbf{k},t-t^{\prime})=-i\left\langle \mathcal{T}
\beta _{\mathbf{k}}(t)\beta _{-\mathbf{k}}(t^{\prime })\right\rangle $,
$D^{++}(\mathbf{k},t-t^{\prime})=-i\left\langle \mathcal{T}
\beta _{-\mathbf{k}}^{\dagger }(t)\beta _{\mathbf{k}}^{\dagger }(t^{\prime })
\right\rangle $, \noindent where $\left\langle \quad \right\rangle$ represents
an average over the ground state. The spin-wave Green's functions satisfy the
 Dyson equations: $D^{\mu \nu }
(\mathbf{k},\omega)=D_{0}^{\mu \nu}(\mathbf{k},\omega )+\sum_{\gamma\delta}
D_{0}^{\mu \gamma }(\mathbf{k},\omega )\Pi ^{\gamma \delta}(\mathbf{k},\omega)
D^{\delta \nu }(\mathbf{k},\omega )$, where $\mu ,\nu =\pm $. The free Green's
functions are $D_{0}^{-+}(\mathbf{k},\omega)=1/(\omega-\omega_{\mathbf{k}}^{0}+
i\eta )$, $D_{0}^{+-}(\mathbf{k},\omega )=1/(-\omega-\omega_{\mathbf{k}}^{0}
+i\eta )$, $D_{0}^{--}(\mathbf{k},\omega)=D_{0}^{++}(\mathbf{k},\omega )=0$,
$(\eta \rightarrow 0^{+})$, and $\Pi^{\gamma \delta }(\mathbf{k},\omega )$ are
the self-energies generated by the interaction between holes and spin waves.
We calculate the spin-wave self-energies in the self-consistent Born
approximation (SCBA), which corresponds to consider only ''bubble'' diagrams
with dressed hole propagators, as illustrated in Figure 1. These diagrams
describe decay of spin waves into ''particle-hole'' pairs. The spin-wave
self-energies can then be written in terms of the hole spectral function,
$\rho(\mathbf{q},\omega)$, as

\begin{equation}
\Pi ^{\gamma \delta }(\mathbf{k},\omega )=\frac{1}{N}\sum_{\mathbf{q}}
U^{\gamma \delta}(\mathbf{k},\mathbf{q})\left[ Y(\mathbf{q},-\mathbf{k};
\omega )+Y(\mathbf{q}-\mathbf{k},\mathbf{k};-\omega )\right] ,  \label{(7)}
\end{equation}
with
\[
Y(\mathbf{q},-\mathbf{k};\omega )=\int_{0}^{+\infty }d\omega ^{\prime
}\int_{-\infty }^{0}d\omega ^{\prime \prime }\frac{\rho (\mathbf{q},
\omega^{\prime })\rho (\mathbf{q}-\mathbf{k},\omega ^{\prime \prime })}
{\omega +\omega ^{\prime \prime }-\omega ^{\prime }+i\eta },
\]
and, $U^{--}(\mathbf{k},\mathbf{q})=U^{++}(\mathbf{k},\mathbf{q})=V(\mathbf{q},
-\mathbf{k})V(\mathbf{q}-\mathbf{k},\mathbf{k})$, $U^{+-}(\mathbf{k},
\mathbf{q})=V(\mathbf{q}-\mathbf{k},\mathbf{k})^{2}$, $U^{-+}(\mathbf{k},
\mathbf{q})=V(\mathbf{q},-\mathbf{k})^{2}$. The relations
$\Pi^{-+}(\mathbf{k},\omega)=\Pi ^{+-}(-\mathbf{k},-\omega )$ and
$\Pi ^{--}(\mathbf{k},\omega )=\Pi ^{++}(\mathbf{k},\omega )$ are verified, the
last implying $D^{--}(\mathbf{k},\omega )=D^{++}(\mathbf{k},\omega )$. The SCBA
provides a spectral function for the holes,$^{9-15}$ that is composed of a
coherent quasi-particle peak and an incoherent continuum, taking the
approximate forms, respectively, $\rho ^{coh}(\mathbf{q},\omega)=a_{0}
\delta(\omega-\varepsilon_{\mathbf{q}})$ with $a_{0}\simeq\left(J/t\right)^
{2/3}$,
and $\rho ^{incoh}(\mathbf{q},\omega )=h\theta (|\omega |-zJ/2)\theta(2zt+zJ/2-
|\omega |)$ with $h\simeq\left( 1-a_{0}\right) /2zt$, the energies are measured
with respect to the Fermi level,  and the
quasi-holes fill up a Fermi surface consisting of pockets, of approximate
radius $q_{F}=\sqrt{\pi \delta }$ , located at momenta $\mathbf{q}_{i}=
(\pm\pi/2,\pm \pi/2)$ in the Brillouin zone, the quasi-particle dispersion
being, near $\mathbf{q}_{i}$, written as $\varepsilon_{\mathbf{q}}=
\varepsilon_{min}+(\mathbf{q}-\mathbf{q}_{i})^{2}/2m$, with an
effective mass $m\simeq 1/J$. The self-energies will then present three
contributions, $\Pi^{\gamma\delta}(\mathbf{k},\omega)=\Pi _{c,c}^{\gamma\delta}
(\mathbf{k},\omega )+\Pi _{c,ic}^{\gamma \delta}(\mathbf{k},\omega )+
\Pi_{ic,ic}^{\gamma\delta}(\mathbf{k},\omega)$, corresponding, respectively,
to transitions of holes within the coherent band, between the coherent and
incoherent bands, and within the incoherent band. We have calculated the
different contributions to lowest order in the hole concentration $\delta $.

The change in the staggered magnetization induced by the interaction
 between holes and spin waves $(6)$, is written in  terms of the spin-wave
 Green\'{}s functions, as

\begin{eqnarray}
\Delta M &=&-\sum_{\mathbf{k}}\frac{2}{(1-\gamma _{\mathbf{k}}^{2})^{1/2}}
\int_{0}^{+\infty }\frac{d\omega }{2\pi }\left[ 2\mathrm{Im}D^{+-}(\mathbf{k},
\omega)\right.\nonumber \\
&&\left. \qquad \qquad \qquad -\gamma _{\mathbf{k}}\mathrm{Im}\left(
D^{++}(\mathbf{k},\omega )+D^{--}(\mathbf{k},\omega )\right) \right] .
\label{(8)}
\end{eqnarray}
To lowest order in the hole concentration $\delta $, $(8)$ gives 

\begin{eqnarray}
\Delta M &=&-\sum_{\mathbf{k}}\frac{2}{(1-\gamma _{\mathbf{k}}^{2})^{1/2}}
\left[-\frac{\gamma_{\mathbf{k}}}{2\omega_{\mathbf{k}}^{0}}
\mathrm{Re}\Pi^{--}(\mathbf{k},\omega _{\mathbf{k}}^{0})\right.  \nonumber \\
&&\left. \quad +\int_{0}^{+\infty }\frac{d\omega }{\pi }\left(
\frac{\mathrm{Im}\Pi ^{-+}(\mathbf{k},\omega )}{(\omega +
\omega _{\mathbf{k}}^{0})^{2}}+\gamma _{\mathbf{k}}
\frac{\mathrm{Im}\Pi ^{--}(\mathbf{k},\omega )}
{\omega^{2}-(\omega _{\mathbf{k}}^{0})^{2}}\right) \right] .  \label{(9)}
\end{eqnarray}
Evaluating $(9)$, one finds that the behavior of
the staggered magnetization is essentially determined by the coherent motion
of holes, and moreover, that it is governed by the
imaginary part of the self-energies, i.e., the contributions

\begin{eqnarray*}
\mathrm{Im}\Pi _{c,c}^{-\pm }(\mathbf{k},\omega ) &=&zJ\sqrt{\delta}
a_{0}^{2}\left(\frac{t}{J}\right)^{2}\frac{1}
{\sqrt{\pi }k(1-\gamma _{\mathbf{k}}^{2})^{1/2}}F^{-\pm }(\mathbf{k},\omega )
\\
&&\times \left[ \sqrt{1-s^{2}(g)}\theta (1-|s(g)|)
-\sqrt{1-s^{2}(-g)}\theta (1-|s(-g)|)\right] ,
\end{eqnarray*}
with
\begin{eqnarray*}
F^{--}(\mathbf{k},\omega )&=&\left[ \cos (k_{x}g)+\cos (k_{y}g)\right]
-\gamma _{\mathbf{k}}\left[ \cos k_{x}\cos (k_{x}g)+\cos k_{y}
\cos(k_{y}g)\right] ,
\end{eqnarray*}
\begin{eqnarray*}
F^{-+}(\mathbf{k},\omega )&=&\left[ (\cos k_{x}-\cos k_{y})
\left(\cos(gk_{x})-\cos (gk_{y})\right)/2 \right.
\\
&&\left. -(1-\gamma _{\mathbf{k}}^{2})^{1/2}\left( \sin k_{x}\sin(gk_{x})
+\sin k_{y}\sin(gk_{y})\right)-2(1-\gamma_{\mathbf{k}}^{2})\right] ,
\end{eqnarray*}
\noindent where $s(g)=(1-g)k/2q_{F}$ and $g=2\omega/Jk^{2}$, while

\[
\mathrm{Re}\Pi _{c,c}^{--}(\mathbf{k},\omega _{\mathbf{k}}^{0})=
-zJ\delta a_{0}^{2}\left(\frac{t}{J}\right)^{2}
\frac{\gamma_{\mathbf{k}}k^{2}}{8(1-\gamma_{\mathbf{k}}^{2}-(k/2)^{4})}
\frac{(\sin^{2}k_{x}+\sin^{2}k_{y})}{(1-\gamma_{\mathbf{k}}^{2})^{1/2}}
. 
\]
Regarding the incoherent contributions,
\begin{eqnarray*}
\lefteqn{\mathrm{Im}\Pi^{-\pm}_{c,ic}(\mathbf{k},\omega)
+\mathrm{Im}\Pi^{-\pm}_{ic,ic}(\mathbf{k},\omega) = 
\mp zJ\sqrt{\delta}(1-a_{0})^{2}\frac{\pi}{32}}
\\
&\times&\left[
\left(\frac{\omega}{4J}-1\right)I_{1}(\omega)
+\left(4\frac{t}{J}+1-\frac{\omega}{4J}\right)I_{2}(\omega)
+4\frac{t}{J}\frac{a_{0}}{(1-a_{0})}I_{3}(\omega)
\right]
\\
&\times&
\left[
\frac{1}{2\sqrt{\pi}}\frac{k^{3}}{(1-\gamma_{\mathbf{k}}^{2})^{1/2}}
\theta(2q_{F}-k)+
\sqrt{\delta}G^{\pm}(\mathbf{k})\frac{(\sin^{2}k_{x}+\sin^{2}k_{y})}
{(1-\gamma_{\mathbf{k}}^{2})^{1/2}}\theta(k-2q_{F})
\right]
,\end{eqnarray*}
with 
\begin{eqnarray*}
&&G^{-}(\mathbf{k})=\gamma_{\mathbf{k}}\hspace{.5cm},\hspace{.5cm}
G^{+}(\mathbf{k})=1+(1-\gamma_{\mathbf{k}}^{2})^{1/2},
\\
&&I_{1}(\omega)=\theta(\omega/4J-1)\theta(2t/J+1-\omega/4J),
\\ 
&&I_{2}(\omega)=\theta(\omega/4J-1-2t/J)\theta(4t/J+1-\omega/4J),
\\
&&I_{3}(\omega)=\theta(2t/J+1/2-\omega/4J)\theta(\omega/4J-1/2),
\end{eqnarray*}
 is one to two orders of magnitude smaller than 
$\mathrm{Im}\Pi^{-\pm}_{c,c}(\mathbf{k},\omega) $, while 
\begin{eqnarray*}
\mathrm{Re}\Pi^{--}_{c,ic}(\mathbf{k},\omega^{0}_{\mathbf{k}})
+\mathrm{Re}\Pi^{--}_{ic,ic}(\mathbf{k},\omega^{0}_{\mathbf{k}})=
zJ\sqrt{\delta}(1-a_{0})^{2}\frac{t}{J}\frac{1}{4}\left[
\ln 2+\frac{a_{0}}{1-a_{0}}\ln\left(1+4\frac{t}{J}\right)
\right]
\\
\times
\left[
\frac{1}{2\sqrt{\pi}}\frac{k^{3}}{(1-\gamma_{\mathbf{k}}^{2})^{1/2}}
\theta(2q_{F}-k)
+
\sqrt{\delta}\gamma_{\mathbf{k}}
\frac{(\sin^{2}k_{x}+\sin^{2}k_{y})}{(1-\gamma_{\mathbf{k}}^{2})^{1/2}}
\theta(k-2q_{F})
\right]
,\end{eqnarray*}
 is of the same order of magnitude as
$\mathrm{Re}\Pi^{--}_{c,c}(\mathbf{k},\omega)$, though smaller.

As a result, we find that the staggered magnetization $(4)$, calculated with 
$(9)$, is strongly reduced with doping, vanishing at a small hole
concentration, as illustrated in Figure 2. 
 The reduction of the staggered magnetization is generated by
the imaginary part of the self-energies,
$\mathrm{Im}\Pi^{-\pm}$, which imply broadening of
the spin-wave spectral function. The real part of the
self-energy, $\mathrm{Re}\Pi ^{--}$, gives rise to an increase of the staggered
magnetization, which however is one order of magnitude smaller than the
decrease due to the imaginary part of the self-energies. 
 The increase of the staggered magnetization arising from the real
part of the self-energy results from the coherent motion of holes, while the
incoherent motion leads to a decrease, though with a smaller amplitude.
We find a critical hole concentration that for $t/J=3$ is 
$\delta _{c}\simeq 0.07$, whereas for $t/J=4$ is 
$\delta _{c}\simeq 0.05$. The value for $\delta_{c}$, of a few percent,
is consistent with experimental data for the copper oxide
high-$T_{c}$ superconductors. The critical hole concentration 
$\delta _{c}$ is smaller than the
hole concentration leading to the vanishing of the spin wave velocity
(e.g., $\delta_{sw}\simeq 0.23$ for $t/J=3$), or the concentration at 
which all spin waves become
overdamped ($\delta ^{*}\simeq 0.17$ also for $t/J=3$), in the same 
approach.$^{5}$ This is
because the staggered magnetization is specially influenced by the strong
damping effects induced by hole motion. Khaliullin and Horsch$^{8}$ did not
consider damping effects, and 
concluded that the long-range order disappears as a result of the incoherent
motion of holes, however having estimated a decrease of the staggered
magnetization due to the incoherent motion of holes that is over one
order of magnitude larger than the one calculated by us. Gan and Mila$^{7}$
studied the effects of damping on the staggered magnetization, though
considering the scattering of spins by holes, i.e. a four-particle interaction
with "uncondensed" bosons. Our results, giving the vanishing of the
magnetization for a hole concentration where the spin wave
velocity is still finite, suggest that, even when long-range order has
disappeared, strong AF correlations persist, which allow spin wave
excitations to exist, for length scales less than the magnetic correlation
length. This is in fact experimentally observed.$^{1}$

In conclusion, we have shown that the staggered magnetization of a
two-dimensional antiferromagnet is significantly reduced as a function of
doping due to the strong interaction between holes and spin waves. The
motion of holes generates decay of spin waves into ''particle-hole'' pairs,
leading to the destruction of the long-range magnetic order at a small hole
concentration. This effect is mainly determined by the coherent motion of
holes. The calculated critical hole concentration is in agreement with
experimental data for the doped copper oxide high-$T_{c}$ superconductors.

We also note that NMR measurements, reported in Ref. 16, show damping of 
the low-energy spin excitations in the doped $CuO_{2}$ planes due to 
``particle-hole'' excitations, which supports the mechanism for destruction 
of the long-range order presented in this work.

\bigskip

We thank T. Imai for bringing Ref. 16 to our attention.

\bigskip

\noindent {\LARGE References:}

\noindent $^{\dagger }$fdias@alf1.cii.fc.ul.pt

\noindent $^{1}$R.J. Birgeneau and G. Shirane, in \textsl{Physical
Properties of High Temperature Superconductors}, ed. D.M. Ginzberg (World
Scientific, New Jersey, 1990).

\noindent $^{2}$J. Gan, N. Andrei and P. Coleman, J. Phys.: Condens. Matter 
\textbf{3}, 3537 (1991).

\noindent $^{3}$I. R. Pimentel and R.Orbach, Phys. Rev. B \textbf{46}, 2920
(1992).

\noindent $^{4}$K.W. Becker and U. Muschelknautz, Phys. Rev. \textbf{48},
13826 (1993).

\noindent $^{5}$I. R. Pimentel, F. Carvalho Dias, L. M. Martelo  and R.
Orbach (accepted in Phys. Rev. B).

\noindent $^{6}$T. V. Ramakrishnan, Physica B \textbf{163}, 34 (1990).

\noindent $^{7}$J. Gan and F. Mila, Phys. Rev. B \textbf{44}, 12624 (1991).

\noindent $^{8}$G. Khaliulin and P. Horsch, Phys. Rev. B \textbf{47}, 463
(1993).

\noindent $^{9}$C.L. Kane, P.A. Lee, and N. Read, Phys. Rev. B \textbf{39},
6880 (1989).

\noindent $^{10}$F. Marsiglio , A. E. Ruckenstein, S. Schmitt-Rink,
 and C. M. Varma, Phys. Rev. B \textbf{43}, 10882 (1991).

\noindent $^{11}$G. Martinez and P. Horsch, Phys. Rev. B \textbf{44}, 317
(1991).

\noindent $^{12}$Z. Liu and E. Manousakis, Phys. Rev. B \textbf{45}, 2425
(1992).

\noindent $^{13}$E. Dagotto, Rev. Mod. Phys. \textbf{60}, 763 (1994).

\noindent $^{14}$N.M. Plakida, V.S. Oudovenko and V. Yu. Yushanhai, Phys.
Rev. B \textbf{50}, 6431 (1994).

\noindent $^{15}$B. Kyung and S. Mukhin, Phys. Rev. B \textbf{55}, 3886
(1997).

\noindent $^{16}$K.R. Thurber, A.W. Hunt, T. Imai, F.C. Chou, and
 Y.S. Lee, Phys. Rev. Lett. \textbf{79}, 171 (1997).

\newpage

\noindent {\large Figure Captions:}

\medskip

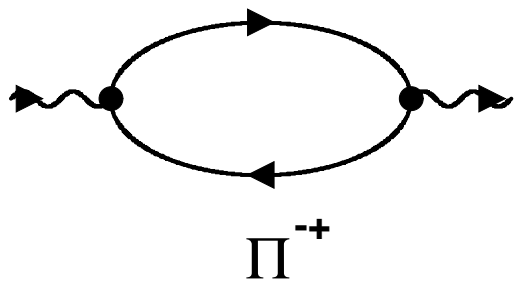
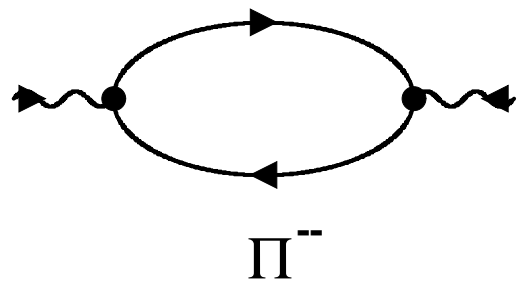
\noindent FIG. 1. Spin-wave self-energies in the SCBA.

\smallskip

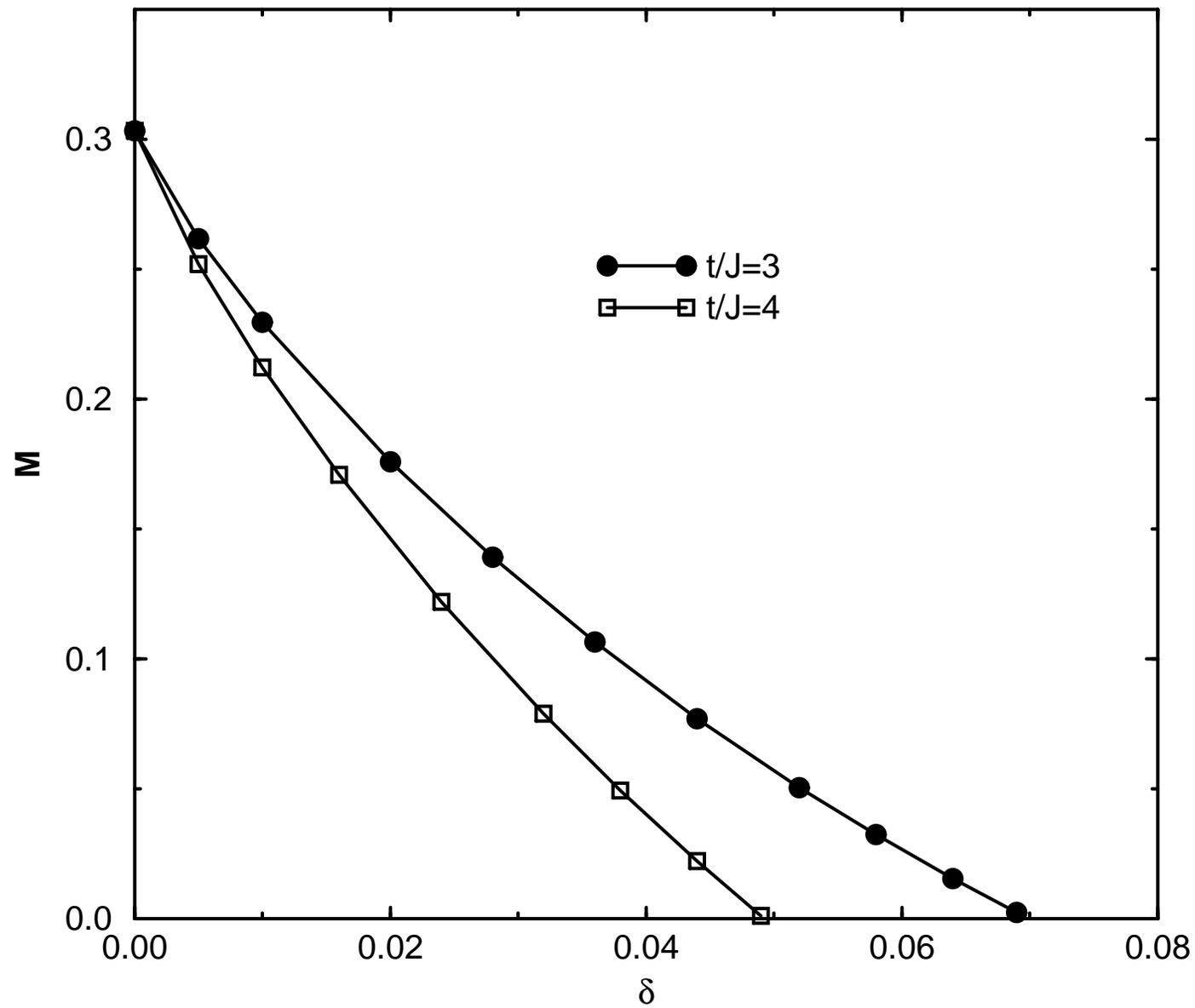
\noindent FIG. 2. The staggered magnetization per spin \textsl{vs} 
hole concentration for different values of t/J.

\end{document}